\documentstyle[12pt]{article}
\title{The stationary KdV hierarchy and $so(2,1)$ as a spectrum
generating algebra} 
\author{H.-D.~Doebner\thanks {E-mail:
    asi@pt.tu-clausthal.de}\\ \small Arnold-Sommerfeld Institute for
  Mathematical Physics, TU Clausthal,\\ \small Leibnizstra\ss e 10,
  38678 Clausthal-Zellerfeld, Germany \and R.Z.~Zhdanov\thanks{E-mail:
    rzhdanov@apmat.freenet.kiev.ua}\\ \small Institute of Mathematics
  of the National Academy of Sciences of Ukraine, \\ \small
  Tereshchenkivska Street 3, 252004 Kyiv, Ukraine }
\date{} \let\ve\varepsilon  
\let\ds\displaystyle  \newtheorem{theo}{Theorem}
\newtheorem{lem}{Lemma}
 \begin{document}

\maketitle
\begin{abstract}
  The family ${\cal F}_L$ of all potentials $V(x)$ for which the
  Hamiltonian $H=-{d^2\over dx^2} + V(x)$ in one space dimension
  possesses a high order Lie symmetry is determined. A sub-family
  ${\cal F}^2_{SGA}$ of ${\cal F}_L$, which contains a class of
  potentials allowing a realization of $so(2,1)$ as spectrum
  generating algebra of $H$ through differential operators of finite
  order, is identified. Furthermore and surprisingly, the families
  ${\cal F}^2_{SGA}$ and ${\cal F}_L$ are shown to be related to the
  stationary KdV hierarchy. Hence, the \lq harmless\rq\ Hamiltonian
  $H$ connects different mathematical objects, high order Lie
  symmetry, realization of $so(2,1)$-spectrum generating algebra
  and families of nonlinear differential equations. We describe in a
  physical context the interplay between these objects.
\end{abstract}

\section*{I. Motivation and Background}

In the two internal reports for the International Center of
Theoretical Physics (ICTP) \cite{doe1,doe1a} written in the early
seventies, a complete classification of symmetric Hamiltonians in one
space dimension on $L^2({\bf R}^1_x,dx)$
\begin{equation}
\label{1}
H=\gamma {d^2 \over dx^2}+V(x),\quad \gamma < 0
\end{equation}
having the Lie algebra $so(2,1)$ as a \lq spectrum generating
algebra\rq\ (SGA) has been obtained. This result has been published
only recently in connection with a Lecture on Memory of A.O. Barut
\cite{doe2}. In \cite{doe1,doe1a} the following definition of SGA is
used: a differential operator $A$ of the order $n'$ has a spectrum
generating (Lie) algebra $L$ with generators $g_i,\; (i=1,...,m,\;
m={\rm dim} \,{L}) $ if there exist a realization $R$ of $ L$ through
differential operators of an order $n \geq n'$, such that
\begin{equation}
   \label{h}
   A=\sum\limits_{i=0}^m \alpha_iR(g_i),
   \quad \alpha_i\in {\bf R}.
\end{equation}

The Hamiltonian (\ref{1}) is a differential operator on $L^2({\bf
  R}^1_x,dx)$ with $n'=2$. A realization $R$ of $so(2,1)$ with
standard basis $\cal L$ spanned by $g_1,g_2,g_3$ through $n$th order
differential operators on a suitable complex function space over $x$
reads
\begin{equation}
   \label{h2}
   R(g_i)=\sum\limits_{j=0}^n a_{ij}(x){d^j\over d
       x^j},\quad i=1,2,3,
\end{equation}
where $a_{ij}(x)$ are complex functions such that the
commutation relations
\begin{equation}
\begin{array}{l}
\label{3}
[R(g_1),\ R(g_2)] = - R(g_3),\quad [R(g_2),\ R(g_3)] = R(g_1),\\[2mm]
[R(g_3),\ R(g_1)] = - R(g_2)
\end{array}
\end{equation}
are fulfilled. So $R(so(2,1))$ is a SGA of (\ref{1}) if there exist constants
$\alpha_i\in{\bf R},\quad (i=1,2,3)$ such that the equality
  \begin{equation}
   \label{h4}
   H\equiv \gamma {d^2\over d^2 x} + V(x)=
   \sum\limits_{j=1}^{3}\alpha_j R(g_j)
 \end{equation}
holds.

The relations (\ref{h2})--(\ref{h4}) impose restrictions both on the
coefficients $a_{ij}(x)$ and on the potential $V(x)$. Solving these
restrictions we find two different families ${\cal
  F}_{SGA}=\left\{{\cal{F}}^{(2)}_1,\ {\cal{F}}^{(n)}_2,\ n\ge
    2\right\}$ of those potentials which allow $so(2,1)$ as SGA for
  (\ref{1}). Now, we can use properties of the representation of
  $so(2,1)$ through $R(so(2,1))$ in order to calculate the spectrum of
  $H$, if $R(so(2,1))$ acts, e.g., on $L^2({\bf R}^1_x,dx)$. If,
  furthermore, $R(g_i)$ are essentially self-adjoint in a common dense
  domain and if the representation is integrable, then we can use the
  known theory for unitary representations of $so(2,1)$.  This is the
  background of the term \lq spectrum-generating algebra\rq\ as
  suggested in \cite{bar,nee}. There were many results in this field
  for different Hamiltonians, Lie algebras and physical systems (see,
  e.g., the recent review \cite{bom}) but no general study in the
  sense of \cite{doe1,doe1a}.

  The motivation of the present paper is to show (in Section III) that
  ${\cal{F}}^{(n)}_2\subset {\cal F}_{SGA}$ can be read from the
  stationary KdV hierarchy and that this surprising connection between
  the KdV hierarchy and $so(2,1)$ has its origin in a certain higher
  order Lie symmetry of the corresponding Hamiltonian (\ref{h4}). In
  Section II we sketch the results of \cite{doe1,doe1a} on which our
  discussion are based.

\section*{II. On Some Known Results}

To classify those $V(x)$ which are solutions of (\ref{h2})--(\ref{h4})
we reduce the calculation to special choices of parameters $\alpha_1,
\alpha_2, \alpha_3$ through transformations of the standard basis
$\cal M$ (which is not unique) to another standard basis ${\cal
  {M}}'$. As a result, we reduce the problem to the following two
cases $(\lambda\in {\bf R}, \lambda\not=0)$:
\begin{eqnarray}
   \label{h6}
   &\mbox{Case 1.}& \alpha_1^2 + \alpha_2^2 \not= \alpha_3^2 \quad
   \mbox{with}\quad H=\lambda R(g_i),\;i=1,2,\\ 
   \label{h7}
&\mbox{Case 2.}& \alpha_1^2 +
 \alpha_2^2 = \alpha_3^2 \quad \mbox{with}\quad H=\lambda (R(g_1)+R(g_3)).
\end{eqnarray}
Case 2 is denoted as the \lq light cone case\rq\ . In both cases
(\ref{h2})--(\ref{h4}) lead for a fixed $n$ to set of coupled
differential equations of the order $n$ for $a_{ij}(x),\; (i=1,2,3,
\;j=1,\ldots,n)$ and $V(x)$. We assume that $R(g_i)$ are symmetric
operators and that $V(x)$ is a real function. A clumsy but
straightforward calculations shows that in Case 1 a solution exists
only for $n=2$ with a family ${\cal{F}}^{1}_{SGA} $ of corresponding
potentials
\[
{\cal{F}}^{1}_{SGA}(\lambda_1,\lambda_2,C)=\{V(x)\;|\;V(x) = \lambda_1(x
- C)^2 + \lambda_2(x - C)^{-2},
\]
\begin{equation}
   \label{h8}\lambda_1,\lambda_2, c\in {\bf R},\lambda_1\not=0\}.
\end{equation}

In Case 2 a solution exists for all $n\geq2$. The corresponding family
${\cal{F}}^{2}_{SGA}$ consists of potentials that are solutions of
nonlinear differential equation
\begin{equation}
\label{5b}
{\cal F}^{2}_{SGA}=\left\{V(x)\ |\ -\left (\frac{x}{2} V' + V\right )
  + \sum\limits_{j=0}^{N-1} C_j {F}_j + {F}_N = 0\right\},
\end{equation}
where $N=\left[{n-1\over 2}\right],\ n\ge 2$; $ {F}_i$ are some
polynomials of $V(x)$ and its derivatives up to the order $2i+1$
($i=0,1,\ldots,N$) which will be derived later in another context
(see, below (\ref{r1}), (\ref{r2})); $C_0, C_1,\ldots,C_{N-1}$ are
arbitrary real constants.

The family ${\cal{F}}^{2}_{SGA}$ has a peculiar structure, the facts
that the equations are equal for $n=2N+1$ and $n=2N+2$,\ 
$N=1,2,\ldots$ and the equation for $n=n_1>n_2$ contains all terms of
the equation for $n=n_2$ are two of these peculiarities. This
structure was not elucidated in \cite{doe1,doe1a}\footnote{This was
  the reason why the authors of \cite{doe1,doe1a} decided to present
  the results as internal reports.}. Relations to other mathematical
notions and objects were not found. The present paper fills this gap.

In order to simplify the following calculations we scale the variable
$x$ and thus get $\gamma=-1$. So the Schr\"odinger operator (\ref{1})
takes the form
\begin{equation}
\label{1a}
H=-{d^2 \over dx^2}+V(x).
\end{equation}

\section*{III. SGA, Lie symmetries and KdV hierarchy}

\subsection*{A. Aim and Strategy}

We will show that the family ${\cal F}^2_{SGA}$ (and the the above
mentioned peculiarities) are connected with a high order Lie symmetry
$Q$ of the stationary Schr\"odinger equation
\begin{equation}
   \label{6}
   \left(-{d^2\over dx^2} + V(x)\right )\psi(x) = 0
\end{equation}
with $[Q, H]=\kappa H$ and that the coefficients $ F_i,\ 
(i=0,1,\ldots, N)$ of equation (\ref{5b}) appear, surprisingly, in the
stationary KdV hierarchy
\[
\sum\limits_{j=0}^{N-1} C_j
{F}_j + {F}_N = 0.
\]
We remind that the stationary KdV hierarchy is obtained successively
by the repeated action of the integro-differential operator (second
recursive operator for the KdV equation \cite{ibr}--\cite{zah})
\begin{equation}
\label{r1}
{\cal R} = -\frac{1}{4}{d^2\over dx^2} - V(x)-\frac{1}{2}V'(x)\left
  ({d\over dx}\right )^{-1}
\end{equation}
generating $F_i, (i=0,1,\ldots)$ through
\begin{equation}
\label{r2}
F_{i+1}={\cal R}\, F_i,\quad i=0,1,\ldots\quad {\rm with}\ 
F_0=-\frac{1}{2}V'(x).
\end{equation}

To get $so(2,1)$ as SGA for $H$, the Lie symmetry has to be specified
through $\kappa=-1$.

Our strategy is the following. We construct at the first step the
family of all potentials $V(x)$ for which $H=-\frac{d^2}{dx^2} + V(x)$
commutes with an $n$th order differential operator $Q\in {\cal L}$,
i.e., the family of potentials with an $n$th order Lie symmetry
provided $\kappa=0$ (Theorem 1). At the second step, we generalize
this result for an arbitrary $\kappa$ (Theorem 2), hence we get the
special case $\kappa=-1$ yielding SGA for the Schr\"odinger operator.

\subsection*{B. Lie symmetries}

First, we remind that an $n$-th order differential operator
\begin{equation}
   \label{7}
   Q = \sum\limits_{j=0}^{n}\ q_j(x) {d^j\over dx^j},\quad {d^0\over
     dx^0}\stackrel{\rm def}{=}1
\end{equation}
is called an $n$-th order (Lie) symmetry operator of the Schr\"odinger
equation (\ref{6}) if it transforms the set of its solutions into
itself. An equivalent (and more algorithmic) definition is the
following one \cite{fun1}. The operator $Q$ is a symmetry operator of
equation (\ref{6}) if there is such an $m$-th order differential
operator $P$ that
\begin{equation}
   \label{8a}
   [Q,\ H] = P H.
\end{equation}

Evidently, if $Q$ is a symmetry operator of an equation $H\psi =0$,
then an operator
\[
 \tilde Q = Q + \sum\limits_{j=0}^N\, \gamma_j H^j
\]
with arbitrary constant $\gamma_j$ is also a symmetry operator. Given
a symmetry operator $Q$, the operator $\tilde Q$ gives no new
information on the structure of a set of solutions of the equation
under study. That is why, it is excluded from further considerations.
In addition, we impose on the symmetry operators in (\ref{8a}), as an
additional constraint, the condition $P=\kappa$ {\boldmath$ 1$}, where
${\kappa}$ is a real constant.

Hence, it is only necessary to consider representatives of the
equivalence classes of the quotient of the linear space of
differential operators (\ref{7}) satisfying the equation
\begin{equation}
 \label{8b}
   [Q,\ H]=\kappa H,\quad \kappa = {\rm const}
\end{equation}
with respect to the equivalence relation
\[
 Q_1\sim Q_2\qquad {\rm if}\qquad Q_1 - Q_2 = \sum\limits_{j=0}^N\,
 \gamma_j H^j
\]
with some $N\in {\bf N}$ and constant $\gamma_j$. We denote this
quotient space as $\cal L$.

To apply this notion to our problem we note that in the light cone
case 2 due to the commutation relations of the algebra $so(2,1)$ the
following equality
\begin{equation}
 \label{com}
   [R(g_2),\ H]=- H
\end{equation}
is fulfilled. Hence $R(g_2)=Q$ is a high order Lie symmetry of
equation (\ref{6}) for the light cone case with $P=-1$, i.e. with
$\kappa=-1$. We add that in the regular case (\ref{h6}) the algebra
$so(2,1)$ does not contain a symmetry operator of the corresponding
Schr\"odinger equation.

Thus a high-order Lie symmetry of the Schr\"odinger equation (\ref{6})
is \lq responsible\rq\ for restricting the potential $V(x)$.
Furthermore, a spectrum generating algebra $so(2,1)$ for $H$ in
(\ref{1}) can be constructed.

Before formulating the principal assertions we prove an auxiliary
lemma.

\begin{lem} The Hamiltonian $H=-{d^2\over dx^2} + V(x)$ commutes with
  an $n$-th order differential operator $Q\in {\cal L}$ (\ref{7}) if
  and only if the Schr\"odinger equation
\begin{equation}
   \label{13}
   \left(-{d^2\over dx^2} + V(x) + \ve\right)\psi=0,\quad
\ve \ne 0
\end{equation}
admits a Lie symmetry of the form
\begin{equation}
   \label{15}
 \hat Q = a(x,\ve){d\over dx} + b(x,\ve)\equiv
   \left ( \sum\limits_{j=0}^N\,a_j(x)\ve^j \right){d\over dx} +
     \sum\limits_{j=0}^Nb_j(x)\ve^j,
\end{equation}
where $N=\left[\frac{n-1}{2}\right],\ a_N=1$ and $\ve$ is a continuous
real parameter.
\end{lem}
We give a sketch of the proof omitting technical details.  To show
that (\ref{13}) implies (\ref{15}) we compute the commutator on the
left-hand side of the equality $[Q,\ H]=0$ and equate coefficients of
the operator ${d^{n+1}\over dx^{n+1}}$. Hence we get in (\ref{7}),
$q_n ={\rm const}$.  Consequently, without loosing generality we may
choose $q_n = 1$ and look for a symmetry operator $Q$ in the form
\begin{equation}
   \label{9}
 Q= {d^n\over dx^n} + \sum\limits_{j=0}^{n-1}\ q_j(x)
 {d^j\over dx^j}.
\end{equation}
Furthermore, as $Q$ belongs to ${\cal L}$, the number $n$
is odd and can be represented as $n=2N+1$ with some $N\in {\bf N}$.

As $Q$ commutes with $H$, it commutes with a shifted Hamiltonian $H +$
$\ve${\boldmath $1$} with an arbitrary constant $\ve$ as well. Thus $Q$
is a symmetry operator of (\ref{13}).

Next, we make use of a well-known fact in the theory of high-order
Lie symmetries of linear differential equations (see, e.g.,
\cite{fun1}).  Let
\[
     X = \sum\limits_{j=0}^n q_{j}(x){d^{j}\over d x_{j}}
\]
be a symmetry operator of the linear differential equation $A\psi = 0$
and $A$, $R$ be differential operators of finite order. Then, $\tilde
X= X + R H$ is also a symmetry operator of the equation $H\psi = 0$.
Choosing the operator $R$ properly we can cancel in $Q$ all powers of
the operator ${d\over dx}$ of the degree $k >1$. According to the last
remark the first-order operator $\hat Q$ obtained in this way is still
a symmetry operator of equation (\ref{13}). Consequently, if the
Schr\"odinger equation (\ref{13}) admits a high order symmetry
operator, then it necessarily admits a first-order Lie symmetry. The
latter can be easily shown to have the form (\ref{15}).

Suppose now that (\ref{13}) admits a first-order Lie symmetry of the
form (\ref{15}).  Then we can cancel all the powers of $\varepsilon$
by adding an appropriately chosen polynomial of $H$ with variable
coefficients. As a result, we get $(2N+1)$th order differential
operator which is still a symmetry of the equation under study.
Moreover, it is straightforward to verify that this operator commutes
with $H$ which is the same as what was to be proved.  $\rhd$

Using this result, we conclude that the problem of description of
$(2N+1)$-th order operators $Q\in \cal L$ commuting with the
Schr\"odinger operator $H$ is equivalent to the study of its usual Lie
symmetry given by (\ref{15}). This remarkable fact connects SGA in the
light cone case to the stationary KdV hierarchy.

\begin{theo} The Hamiltonian $H=-{d^2\over dx^2} + V(x)$ commutes with
  an $n$-th order differential operator $Q\in {\cal L}$ if and only if
  the potential $V(x)$ satisfies of the following families ${\cal
    F}^0_{L}$ of nonlinear differential equations
\begin{equation}
\label{hkdv2}
{\cal F}^0_{L}=\left\{V(x)\ |\ G(V)\equiv\sum\limits_{j=0}^{N-1} C_j
  {F}_j + {F}_N = 0,\ N\in {\bf N}\right\},
\end{equation}
where $F_i$ are polynomials in $V(x)$ and its derivatives forming the
stationary KdV hierarchy (\ref{r2}), $C_0,C_1,\ldots,C_{N-1}$ are some
real constants.
\end{theo}
{\bf Proof}.$\quad$ The proof is simplified substantially if we use
Lemma 1 for $H + \ve$ {\boldmath $1$}, because this gives a possibility
to use the well-known technique of the soliton theory.
 
Inserting (\ref{13}), (\ref{15}) into the invariance condition $[H +
\ve${\boldmath $1$},\ $\hat Q]=R(H+\ve${\boldmath$1$}) we get
\begin{eqnarray*}
  &&\left [-{d^2\over dx^2}+V(x)+\ve\mbox{\boldmath $1$},\ a(x,\ve){d\over
      dx} + b(x,\ve)\right ]\\ &&\quad=r(x,\ve)\left (-{d^2\over dx^2}
    + V(x) + \ve\mbox{\boldmath $1$}\right ).
\end{eqnarray*}
This gives a system of determining equations for the coefficients $a,
b$ which are $N$th order polynomials in $\ve$
 \begin{eqnarray*}
 &&a''(x,\ve) +2 b'(x,\ve)=0,\\
 && b''(x,\ve)+a(x,\ve) V'(x)+2 a'(x,\ve)(V(x)+\ve) = 0,
\end{eqnarray*}
where primes denote differentiation with respect to $x$. Using the
$\ve$-depen\-dence we find after integration the forms of the functions
$b_i(x)$
\begin{equation}
 \label{16}
 b_i(x)=-{\ds \frac{1}{2}} a'_i(x) + B_i,\quad i=0,1,\ldots,N \\
\end{equation}
and $N+2$ recurrence relations for $a_j(x)$ depending on
$V(x)$ and its derivatives
\begin{equation}
\label{17}
\begin{array}{rcl}
  a_N(x)&=&1,\\[2mm] a'_{j-1}(x)&=&-{\ds \frac{1}{4}} a'''_j(x) -
   V(x)a'_j(x) - {\ds \frac{1}{2}} V'(y) a_j(y),
\end{array}
\end{equation}
where $B_j$ are integration constants,\ $a_{j}(x)\stackrel{\rm
  def}{=}0$ for $j=-1$,\ $j=N,N-1,\ldots,0$. The set of equations
(\ref{17}) can be considered as differential equations for $V(x)$.

The first $N+1$ relations of (\ref{17}) are solved by subsequent
integrations yielding the expressions for the functions
$a_0(x),\ldots,a_{N-1}(x)$ via the function $V(x)$ and its
derivatives. Substituting these results into the last equation for
$j=0$ we arrive at an $(2N+1)$th order nonlinear differential equation for
$V(x)$.  

To reveal the structure of (\ref{17}) we introduce new functions
${\cal U}_0(x)$,\ ${\cal U}_1(x)$,$\ldots$ by the following recurrence
relation:
\begin{equation}
 \label{18}
 \begin{array}{l}
   {\cal U}_{j}(x)={\cal P}\, {\cal U}_{j-1},\quad {\cal
     U}_{-1}\equiv 1,\\[2mm] {\cal P}= -{\ds\frac{1}{4}}{\ds{d^2\over
       dx^2}} - V(x) + {\ds\frac{1}{2}}\left ({\ds{d\over dx}}\right
   )^{-1}V'(x),
 \end{array}
\end{equation}
where $j=0,1,\ldots$ and ${\cal U}_{-1}(x) \stackrel{\rm def}{=}1$.
Note that ${\cal P}$ is the first recursive (integro-differential)
operator for the KdV equation (see, e.g. \cite{ibr}--\cite{zah}). The
action of ${\cal P}$ on some initial conserved density ${\cal
  U}_0=-\frac{1}{2}V(x)$ yields the whole hierarchy of the conserved
densities ${\cal U}_1, {\cal U}_2,\ldots $.

The essential point of the proof is that we can solve (\ref{17}) 
for $j=N, N-1,\ldots, 1$ in terms of ${\cal U}_j(x)$
\begin{equation}
 \label{19}
 a_{N-j}(x)={\cal U}_{j-1} + \sum\limits_{k=1}^{j-1}\,C_{N-k}\,
 {\cal U}_{j-k-1}(x) + C_{N-j},\quad j=1,\ldots,N,
\end{equation}
where $C_0,\ldots,C_{N-1}$ are integration constants independent of
constants $B_j$, ($j=0,1,\ldots,$ $N$).

As $a_{-1}(x)\stackrel {\rm def}{=}0$, the equation for $j=0$ can be
rewritten as ($D_x$ denotes $\frac{d}{dx}$)
\[
 D_x\circ {\cal P}\, a_0(x)=0
\]
or, equivalently,
\begin{equation}
 \label{last}
 D_x\circ \left(\sum\limits_{j=0}^{N-1}\,C_{j}\, {\cal P}^{j} + {\cal
     P}^{N}\right) {\cal U}_0=0,
\end{equation}
where ${\cal U}_0 =-\frac{1}{2}V$, as starting point of the recursion.

The solutions of (\ref{last}) generate all the solutions of
(\ref{16}), (\ref{17}) which are differential equations for those
$V(x)$ that allow for $H+\ve${\boldmath $1$} a Lie symmetry with
respect to a first-order operator (\ref{15}).

In what follows we will show that (\ref{last}) coincides with the
stationary higher KdV equation. Using the operator identity
\[
D_x\circ {\cal P}^j\equiv \left (D_x\circ {\cal P}\circ
  D_x^{-1}\right )^j\circ D_x
\]
we relate the integro-differential operator ${\cal P}$ with the
second recursive operator for the KdV equation ${\cal R}$,
\[
 {\cal R}=D_x\circ {\cal P}\circ D_x^{-1}=-\frac{1}{4}D_x^2 - V -
 \frac{1}{2}V'D_x^{-1}.
\]

With the operator ${\cal R}$ we can represent equation (\ref{last}) in
the following form:
\[
 \left(\sum\limits_{j=0}^{N-1}\, C_j {\cal R}^{j} +
 {\cal R}^{N}\right)\circ D_x\, {\cal U}_0 = 0,\quad
{\cal U}_0= -\frac{1}{2} V'(x).
\]

Taking into account that $D_x\, {\cal U}_0 = -\frac{1}{2}V' = {F}_0$
and that $F_j={\cal R}^jF_0$ we get finally (\ref{hkdv2}). Thus the
Schr\"odinger equation (\ref{13}) admits a Lie symmetry of the form
(\ref{15}) if and only if the potential $V(x)$ is a solution of
equation (\ref{hkdv2}). $\rhd$

\begin{theo} The Hamiltonian $H=-{d^2\over dx^2} + V(x)$ admits an $n$-th 
  symmetry operator $Q\in {\cal L}$ with $[Q, H] =\kappa H$ if and only
  if the function $V(x)$ satisfies nonlinear differential equation
\begin{equation}
\label{diff}  
\kappa\left (\frac{x}{2} V' + V\right ) +
  \sum\limits_{j=0}^{N-1} C_j  {F}_j +  {F}_N = 0
\end{equation}
with $N=[\frac{n-1}{2}],\ n\ge 1$ and with ${F}_j={\cal R}^jF_0$ under
${\cal F}_0 = -\frac{1}{2}V'$; $C_i$ are some constants.
\end{theo}

\noindent
{\bf Proof}.$\quad$ Computing the commutator on the left-hand side of
(\ref{8b}) and equating coefficients of the operators ${d^{n+1}\over
  dx^{n+1}}$ we have $q_n ={\rm const}$. Consequently, without loosing
generality we may choose $q_n = 1$ and look for a symmetry operator
$Q$ in the form (\ref{9}). Furthermore, as the operator $Q$ belongs to
${\cal L}$, $n$ is odd and, consequently, can be represented as
$n=2N+1$ with some $N\in {\bf N}$.

With this remark (\ref{8b}) reads as
\begin{equation}
 \label{10}
 \begin{array}{l}
   - {\ds \sum\limits_{j=0}^{n-1}}\, \left (\begin{array}{c} {j}\\{n}
     \end{array}\right )\, V^{(n-j)}\, {\ds { d^j\over dx^j}} -
   {\ds \sum\limits_{i=1}^{n-1} \sum\limits_{j=0}^{i-1}}\, q_i\, \left
     (\begin{array}{c} j\\ i \end{array}\right )\, V^{(i-j)}\, {\ds{
       d^j\over dx^j}} \\[4mm]
      \phantom{- {\ds
       \sum\limits_{j=0}^{n-1}}} - {\ds \sum\limits_{j=0}^{n-1}}\,
   (2q'_j{\ds {d\over dx}} + q''_j)\,{\ds {d^j\over dx^j}} = \kappa
   \left ({\ds {d^2\over dx^2}} - V(x)\right ).
 \end{array}
\end{equation}

Comparing coefficients in front of the linearly-independent operators
${d^j\over dx^j}$,\ $(j=1,\ldots n)$ yields $n$ recurrence
integro-differential relations for the coefficients
$q_i(x,V(x),\kappa)$, ($i=0,1,\ldots, n-1$) in the operator $Q$
\begin{eqnarray}
   q_{n-1}(x)&=&C_{n-1}, \nonumber\\
   q_{j-1}(x)&=& -\frac{1}{2}\Biggl (q_j'(x) +
   \left (\begin{array}{c} j\\ n \end{array}\right )
   V^{(n-j-1)}(x) + \sum\limits_{i=j+1}^{n-1}\,
   \left (\begin{array}{c} j\\ i \end{array}\right )\label{11}\\
    &&\times \int\limits^{x} q_i(y) V^{(i-j)}(y) dy + \kappa \delta_{j2} x \Biggr
   ) + C_{j-1},
   \nonumber
\end{eqnarray}
where
\[
 C_j=\left \{\begin{array}{ll} \tilde C_k,& j=2k+1,\\[2mm]
                               0,  & j=2k,\end{array}\right.
\]
${\tilde C}_k$ are arbitrary constants.

Collecting the terms without derivative ${d\over dx}$ in
(\ref{10}) we get an equation for $V(x)$ of the type
\begin{equation}
   \label{12}
   G(V,\kappa) \equiv q_0'' + V^{(n)} + \sum\limits_{j=1}^{n-1}q_j
   V^{(j)} + \kappa V = 0.
\end{equation}
Now we apply Theorem 1. For $\kappa =0$ we know that the equation for $V(x)$
is given by (\ref{hkdv2}). Hence
\begin{equation}
 \label{00}
 G(V,0)\equiv G(V)=\sum\limits_{j=0}^{N-1}\, C_j {F}_{j} +
 {F}_{N},\quad N\in {\bf N}
\end{equation}
holds with some constant $C_0,C_1,\ldots,C_{N-1}$.

On the other hand, an analysis of relations (\ref{11}), where $\kappa$
appears for $j=2$ only, and (\ref{12}) yields that $V(x)\in {\cal
  F}^{\kappa}_L$, where
\begin{equation}
 \label{14}
 {\cal F}^{\kappa}_{L}=\left\{V(x)\ |\ G(V,\kappa) \equiv G(V) + \kappa
   \left (\frac{x}{2} V' + V\right)=0\right\}.
\end{equation}
Combining the relations (\ref{00}) and (\ref{14}) we arrive at
(\ref{diff}). $\rhd$

This concludes the discussion of high order Lie symmetry of the
Hamiltonian $H=-\frac{d^2}{dx^2}+V(x)$.

\subsection*{C. Relation to spectrum generating algebras}

The Lie symmetry $[Q, H]=\kappa H$ is related to the spectrum
generated algebra $so(2,1)$ of $H$ through (\ref{com}), where
$\kappa=-1$.  The realization $R(g_2)$ of $g_2\in so(2,1)$ is given by
$Q$ of the form (\ref{7}), (\ref{11}) through solutions of
(\ref{diff}) with $\kappa=-1$. As $R(g_2)$ is explicitly known, we can
insert $R(g_2)$ and the differential operators $R(g_1)$, $R(g_3)$ into
the commutation relations of the algebra $so(2,1)$ and thus find the
latter (for further details, see \cite{doe1,doe1a}).

With the results obtained in III.B we can elucidate the peculiar
features of the nonlinear differential equation (\ref{5b}) mentioned
in Section II.  The fact that the potentials $V(x)$ are identical for
$n=2N+1$ and $n=2N+2$ is explained as follows. The coefficient in
front of the highest power of the symmetry operator $Q$ is equal to
$1$. Utilizing this property we can cancel this coefficient in the
$(2N+2)$th order symmetry operator $Q$ by subtracting from it the
trivial symmetry $H^{N+1}$ thus getting a $(2N+1)$th order symmetry
operator $\tilde Q$.  Evidently, the latter is admitted by the same
Schr\"odinger equation which is the same as what was claimed. The
stronger statement that the equations for $V(x)$ are identical for
$n=4k, 4k+1, 4k+2, 4k+3,\ k=1,2,\ldots$ is valid for SGA appearing in
\cite{doe1,doe1a}.  The way of constructing of the equation for $V(x)$
used while proving Lemma 1, makes it also evident, why this equation
with some fixed $n=n_1$ contains all the terms of an equation for
$V(x)$ under $n=n_2<n_1$.  Indeed, the equation for $n=n_1$ is
obtained from one for $n_1-1$ with the action of the recursive
operator $\cal R$ and the latter transforms a term $F_i$ into
$F_{i+1}$ and, what is more, ${\cal R}\, 0=$const.

\subsection*{D. Integrability}
Hamiltonians (\ref{1a}) admitting $so(2,1)$ spectrum generating
algebra have a further useful property, they are {\em integrable} in
the sense that the corresponding Schr\"odinger equation (\ref{6}) can
be integrated by quadratures. This is so because (\ref{6}) admits a
first-order Lie symmetry of the form $ X=\xi(x){d\over dx}+\eta(x)$
with
\begin{equation}
\label{lies}
 -\eta''+2V\xi'+V'\xi=0,\quad
 2\eta'+\xi''=0.
\end{equation}

Hence we can apply for integration of equation (\ref{6}) the classical
method (see, e.g., \cite{car}) based on its Lie symmetry. The first integral
for system (\ref{lies}) is given by the following formula:
\[
 -\eta'\xi+V\xi^2-\frac{1}{4}=\alpha\equiv {\rm const}.
\]
Depending on the sign of $\alpha$ the general solution of the equation
(\ref{6}) reads
\[
 \psi(x)=\sqrt{\xi(x)}\left\{\begin{array}{ll}
             C_1f(x)+C_2,&\alpha=0 \\
             C_1\cos a f(x)+C_2\sin a f(x),& \alpha=a^2>0,\\
             C_1\cosh a f(x) +C_2\sinh a f(x),& \alpha=-a^2<0,
             \end{array}\right.
\]
where
\[
 f(x)=\int{dx\over \xi(x)}.
\]
Now inserting the explicit expressions for $\xi(x), \eta(x)$ into the
above formulae yields the general solution of (\ref{6}) provided the
function $V(x)$ fulfill an equation of the form (\ref{5b}).

\section*{IV. Concluding Remarks}

Given a physical observable quantized through linear differential
operator $A$ in $L^2({\bf R}^d, dx^d)$, e.g., the Hamiltonian $H$ as
in (\ref{1}) and $d=1$, a spectrum generating algebra for $A$ is
specified through a Lie algebra $L$ having the dimension $d$ (say,
$so(2,1)$) with generators $g_i$, a realization $R$ of $L$ through
differential operators of the order $n$ and $H=\sum_{j=1}^p \alpha_j
R(g_j)$. A high order Lie symmetry for the linear operator $A$ is
defined through finite order differential operators $Q, P$ with $[Q,
A]=PA$, e.g., $A=H$ and $P=\kappa$. Both SGA and high order Lie symmetry
are different methods with different mathematical structures.  They
model a symmetry of $A$. We have shown that for the Hamiltonian
(\ref{1}) $so(2,1)$-SGA and the Lie symmetry with $\kappa=-1$ are
directly related via (\ref{com}). However, Lie symmetry is more general
than $so(2,1)$-SGA symmetry (see Section III.C). The interesting
result is a connection to the stationary KdV hierarchy. It is
understandable that for the singular case of $so(2,1)$, which reflects
the light cone case, a family of non-linear differential equations for
$V(x)$ appears. But it is, as we already mentioned, surprising that
this family is the stationary KdV hierarchy, a mathematical object not
connected directly to a symmetry concept of observables. We suspect
that the KdV hierarchy is somehow encoded in the geometry of $so(2,1)$
and its realizations. An investigation of Hamiltonians of the type
(\ref{1}) with $d>1$, their Lie symmetry in the above sense and
$L$-SGA with non-compact $L$, $p>3$, seems to be appropriate.

\section*{Acknowledgments} 

One of the author (R.Zh.) gratefully acknowledges a support from the
Alexander von Humboldt Foundation and Technical University of
Clausthal.


\begin{thebibliography}{100}

\bibitem{doe1} H.-D.Doebner and Pirrung B., Spectrum generating
  algebras and canonical realizations, {\em International Center for
    Theoretical Physics, Trieste Report IC/72/77}, 1972.

\bibitem{doe1a} H.-D.Doebner and Pirrung B., A new class of
  Hamiltonians with $so(2,1)$ as spectrum generating algebra, {\em
    International Center for Theoretical Physics, Trieste Report
    IC/75/112}, 1972.

\bibitem{doe2} H.-D.Doebner and H.-J.Mann, in: {\em Proceedings of the
    VIth Wigner Symposium}, Eds.: N.M.Atakashiev, T.H.Seligman,
  K.B.Wolf, World Scientific, 1996, pp.11--23.

\bibitem{bar} Barut A. and B\"ohm A., {\em Phys. Rev. B}, 1965, {\bf
    139}, 1107--1112.

\bibitem{nee} Dotham Y., Gell-Mann M. and Ne'eman Y., {\em Phys.
    Lett.}, 1965, {\bf 17}, 48--51. 

\bibitem{bom} Barut A.O., Bohm A. and Ne'eman Y., {\em Dynamical
    Groups and Spectrum Generating Algebras}, World Scientific,
  Singapure (1988).

\bibitem{ibr} Ibragimov N.Kh., {\em Transformation Groups Applied to
    Mathematical Physics}, Reidel, Dordrecht (1985).

\bibitem{abl} Ablovitz M.J., Segur H. {\em Solitons and the Inverse
    Scattering Transform}, SIAM, Philadelphia (1981).

\bibitem{zah} Zakharov V.E., Manakov S.V., Novikov S.P. and Pitaevski
  L.P. {\em Theory of Solitons: the Inverse Scattering Method},
  Consultants Bureau, New York (1980)

\bibitem{fun1} Fushchych W.I.  and Nikitin A.G., {\sl Symmetries of
    Maxwell's Equations}, Dordrecht, Reidel, 1987.

\bibitem{olv2} Olver P.J., {\em Applications of Lie Groups to
    Differential Equations}, Springer, New York (1986).

\bibitem{car} Cartan E. {\sl Les Syst\'emes Differentiels Ext\`erieurs
    et Leur Applications G\'eometriques}, Paris, Hermann, 1945.

\end{thebibliography}
\end{document}